# Oxygen deficiency induced suppression of JT-distortion and stabilization of charge-ordering in $La_{0.2}Sr_{0.8}MnO_{3-\delta}$

Aga Shahee, R.J.Choudhary, R.Rawat and N. P. Lalla*
UGC-DAE Consortium for Scientific Research, University campus, Khandwa road Indore, India- 452017

## Abstract

Structural phase transition studies employing low-temperature transmission electron microscopy, and low-temperature x-ray diffraction have been carried out on nearly stoichiometric ($\delta$=0.01) and off-stoichiometric ($\delta$=0.12) versions of $La_{0.2}Sr_{0.8}MnO_{3-\delta}$ manganite. The nearly stoichiometric $La_{0.2}Sr_{0.8}MnO_{3-\delta}$ under goes a cubic to Jahn-Teller distorted C-type antiferromagnetic tetragonal phase-transition at 260K. For off-stoichiometric $La_{0.2}Sr_{0.8}MnO_{3-\delta}$ the Jahn-Teller distorted tetragonal phase transition gets totally suppressed and the basic perovskite lattice remains cubic but shows a charge ordered phase. Stabilization of charge ordered phase in the absence of cooperative Jahn-Teller distortion has been attributed to coulomb-repulsion and Hunds coupling energy. The off-stoichiometric sample shows characteristically different physical properties. This has been realized through transport, magnetic and calorimetric measurements. A smooth cross-over from variable range hopping transport to power-law dependence of resistivity ($\rho$) on temperature (T) i.e. $\rho = C.T^{-\alpha}$ has been realized. This has been attributed to multistep inelastic tunneling through channels involving localized states around oxygen vacancy sites.

*Corresponding author: N. P. Lalla (nplallaiuc82@gmail.com)

# INTRODUCTION

Strongly correlated manganites exhibit the phenomena of charge-ordering (CO) [1,2], which depends on parameters like electron-electron coulomb repulsion (U), electron bandwidth (W), intra-atomic Hunds coupling energy ($J_H$) and electron-lattice interaction through Jahn-Teller (JT) distortion ($E_{JT}$) [3]. CO occurs as commensurate and incommensurate modulations and was initially modeled as decorations of $Mn^{+3}/Mn^{+4}$ strips locked with the basic lattice [4,5]. But later a general picture of CO emerged as charge density wave [6,7] and Zner-polaron ordering [8, 9], in which Zener double-exchange (DE) [10] induced delocalization of $e_g$-electrons between Mn-O-Mn bonds gives rise to $Mn^{+3.5+\psi}/Mn^{+3.5-\psi}$ like charge-disproportionation ($\psi$). Experimental [11] and theoretical [12] studies have now verified that, unlike the expectation from a true ionic picture of CO, the actual charge-disproportionation in these systems is very small. The value of $\psi$ ranges from 0.05-0.15 only. The practical realization of CO phase is accompanied by ordering of real orbitals [13] resulting in cooperative JT-distortion (CJTD) [14, 15]. Thus usually charge and orbital ordering (COO) and CJTD should appear simultaneously. But theories based purely on kinetic considerations and band approach does predict the possibility of CE-type charge ordered phase even in the limit of vanishing JT-distortion [16, 17]. The semi-covalence approach of Goodenough [18] also appears to predict the stabilization of COO phase purely based on magnetic interactions. The possible occurrence of such a COO phase has been pointed out as an important issue in the field of manganites [19]. The development of concepts like complex orbital-ordering [20-22] provides further theoretical background for the possibility of a COO phase without JT distortion. But yet such a COO phase has not been practically realized. Study focusing on the possible occurrence of such COO phase is important because it is supposed [19] to throw light on the physics of colossal magnetoresistance. The definition of the problem itself indicates that to investigate the experimental possibility of such a COO phase one should study manganite system in which JT distortion is either naturally too low or its presence can be suppressed. $La_{1-x}Sr_xMnO_3$ manganite appears to be suitable system for such study.

A general phase-diagram of manganite [1,2] indicates that 3D $La_{1-x}Sr_xMnO_3$ has the widest one-electron band-width (W) for higher (x). Aken et al [15] in the case of low doping level have shown that with increasing x the Q2 JT distortion mode gets suppressed leading to insulator to metal transition at x~0.2. At higher x i.e. for x>0.5 $La_{1-x}Sr_xMnO_3$ is either cubic or tetragonal with Mn-O-Mn bond angle equal to $180^o$ and therefore hopping of $e_g$-electrons across it will be the maximum and electron band-width will be the widest. Unlike narrow band manganites like $La_{1-x}Ca_xMnO_3$ and $Pr_{1-x}Ca_xMnO_3$ no CO is expected for this wide band $La_{1-x}Sr_xMnO_3$. In the range of x = 0-0.35 $La_{1-x}Sr_xMnO_3$ has been studied for its interesting structural, transport, magnetization and calorimetric behaviour [23-26]. At higher x values $La_{0.2}Sr_{0.8}MnO_3$ show a paramagnetic cubic to directly C-type antiferromagnetic (AFM) tetragonal phase transition around 260K with a F-type orbital-ordering [27,28]. Together with this, suppression of metallicity and ferromagnetism [29,30] in this wide band manganite [2,31] $La_{1-x}Sr_xMnO_3$ with x=0.5-0.8 [32-34] has also been reported. Here it should be emphasized that the occurrence of COO in a La-Sr-Mn-O system has mainly been studied in its layered perovskite version [35-40]. The report of COO in 3D La-Sr-Mn-O perovskite version is rather sparse [41]. Only recently Bindu et al [41] has shown the occurrence of intermediate CO phase in $La_{0.2}Sr_{0.8}MnO_3$ manganite. But the reason of its stabilization in $La_{0.2}Sr_{0.8}MnO_3$ is unclear. They have shown that CO phase appears at around 265K and then slowly transforms to JT distorted C-type AFM tetragonal phase, which competes with the CO phase. At ~100K C-type AFM tetragonal phase remains as the major phase (~90%). This observation appears to indicate that if the JT distortion, which is responsible for the stabilization of F-type orbital ordered C-type AFM tetragonal phase, is somehow suppressed in $La_{0.2}Sr_{0.8}MnO_3$, CO phase may be retained down to much lower temperatures.

Keeping the above in view we have prepared $La_{0.2}Sr_{0.8}MnO_{3-\delta}$ samples with stoichiometric (or nearly stoichiometric) and off-stoichiometric oxygen contents and have carried out low-temperature structural phase-transition, electrical-transport, calorimetric and magnetization studies on them. These studies have clearly revealed that in off-stoichiometric $La_{0.2}Sr_{0.8}MnO_{3-\delta}$ sample CO phase stabilizes without any type of JT distortion. That is what has been theoretically expected [13-15].

## EXPERIMENTAL

$La_{0.2}Sr_{0.8}MnO_{3-\delta}$ was prepared through conventional solid-state reaction rout using 99.99% pure $La_2O_3$, $SrCO_3$ and $MnO_2$. Stoichiometric mixture of the ingredients was thoroughly mixed using mortar pastel for effectively about 24 hrs and then calcined in air at $1100^oC$ for 24hrs. The calcined powder was reground and fired at $1400^oC$ for 24hrs. The fired powder was reground and pelletized in the form of 14mm x 1mm disks and then finally sintered at $1400^oC$ for 48hrs in air. The furnace was then slowly cooled to room temperature (RT) with a rate of $2^oC$/minute. Few of the final sintered pellets were further annealed at $1000^oC$ for 5 hours under argon atmosphere to make them oxygen deficient. These samples were then subjected to phase-purity, compositional homogeneity and oxygen-stoichiometry characterizations using powder x-ray diffraction (XRD), energy dispersive analysis of x-ray (EDAX) and idiometric titration. The XRD was carried out using a Rigaku (Japan) make diffractometer (model D-max) with angular resolution ($\Delta\theta$) of $0.047^o$ ($8.33 \times 10^{-4}$ rads.). This was equipped with a graphite (002) monochromator and a $LN_2$ based cryostat for sample temperature control below room temperature. The diffractometer was mounted on a rotating anode x-ray generator producing $CuK_\alpha$ x-rays at 11kW. The room temperature XRD data of both the samples are shown in Fig.1. The XRD data were Rietveld refined using space-group Pm-3m. All the peaks were accounted by the Rietveld refinement fit. This confirms that the samples are pure cubic phase. For compositional homogeneity analysis EDAX was carried out with a ~20nm probe at various close by (~50nm) points in a single grain. The variation in the atomic ratio of the elements La, Sr and Mn at different points of the grain was found to match within ±1% of the atomic percent to the synthesized ratio of 10%, 40% &50% respectively. This is well within the typical error of EDAX technique. Idiometric titration revealed the large off-stoichiometry of oxygen in the argon-annealed $La_{0.2}Sr_{0.8}MnO_{3-\delta}$, $\delta=0.12$. For air-sintered sample the oxygen off-stoichiometry $\delta$ was found to be very low, only 0.01. Here after the $La_{0.2}Sr_{0.8}MnO_{3-\delta}$ sample with $\delta=0.01$ will be termed as nearly stoichiometric and that of with $\delta=0.12$ will be termed as off-stoichiometric. The samples were then subjected to structural phase-transition studies, employing low-temperature XRD (LT-XRD) (down to 80K) and low-temperature transmission electron microscopy (LT-TEM) (down to 100K). Low-temperature electrical-transport, magnetization and calorimetric measurements down to ~40K were also carried out. For TEM studies thin samples were prepared using conventional method of ion-beam polishing [42]. For LT-TEM studies a GATAN make liquid-nitrogen based sample holder 636MA was used. Resistivity ($\rho$)-vs-temperature ($\rho$-T) measurements were done using Van der Pauw 4-probe method. For this Keithley make nano-voltmeter model 182, constant-current source model 2400 and Lakeshore make temperature controller model DRC-93CA were used. Zero-field cool (ZFC) and field-cool (FC) magnetization-vs-temperature (M-T) at 200Oe and 5T fields were done using SQUID-VSM (Quantum Design). Heat-capacity measurements were done using a well calibrated relaxation calorimeter.

## RESULTS

The data of LT-TEM observations are shown in Figs.2 and Fig.3. Fig.2(a) and (b) show electron micrographs of nearly stoichiometric $La_{0.2}Sr_{0.8}MnO_{3-\delta}$ taken at room temperature and 100K respectively. It can be seen that the nearly stoichiometric sample posses a smooth and contrast-less microstructure at room-temperature which changes to nearly periodically arranged nano band-like (10-20 nm) contrast at 100K. Fig.3c exhibits magnified view of the nano bands. Figs.2 (d) and (e) show selected area diffraction (SAD) patterns taken along [100] zone at room temperature (i.e. above $T_s$) and at 100K (i.e. below $T_s$) respectively. The SAD in Fig.2 (d) corresponds to the cubic-phase at room temperature, whereas the SAD in Fig.2 (e), which was taken from nano-band region is analyzed to be due to twined nano-bands of the tetragonal phase appearing out of cubic-tetragonal phase transformation. Fig.2(e) corresponds to a composite pattern arising from two type of twin variants of the tetragonal phase, which alternately occur having mirror relation parallel to {011} planes. Due to occurrence of tetragonality (c/a>1) the cubic reflections split in these composite SAD pattern taken along [100] zone. Here we would like to emphasize that unlike the report by Bindu et. al. [41], we could not observe any intermediate CO phase during low-temperature TEM exploration of the nearly stoichiometric $La_{0.2}Sr_{0.8}MnO_{3-\delta}$ sample. To be sure enough about the absence of CO phase we applied the technique of convergent beam electron diffraction (CBED) and searched for higher order Lave zone (HOLZ) rings with radius smaller than that of the (100) HOLZ ring. Presence of extra reciprocal lattice points appearing due to CO super-lattice modulation would create extra reciprocal lattice planes in between the (100) HOLZ and zero order Lave zone (ZOLZ). The presence of these extra reciprocal lattice planes will be reflected as HOLZ rings with radius smaller than the fundamental (100) HOLZ ring. If at all observed, the radius of these HOLZ rings could have been well correlated with the CO modulation period. But even after a rigorous and thorough TEM investigation we could not observe any such HOLZ ring for the nearly stoichiometric sample. The off-stoichiometric ($\delta$=0.12) sample does not show cubic to tetragonal phase transformation on cooling. A typical [100] zone high-resolution (HR) micrograph and corresponding SAD pattern recorded at 100K is shown in Fig.3 (a) and (b) respectively. Fig.3(c) shows temperature scan of [100] zone SAD pattern of the off-stoichiometric $La_{0.2}Sr_{0.8}MnO_{3-\delta}$. This depicts how the diffuse and weakly intense cross-streaks around the main perovskite spots at room temperature, slowly develop into CO super-lattice spots at low-temperatures. These superlattice reflections were found to correspond to a ~9x9 CO modulation [4-6] of the basic perovskite lattice along [011] and [0-11]. The HR micrograph in Fig.3 (a) clearly shows that unlike the typical 1-dimensinal (1D) CE-type charge-ordering occurring along [011] direction in narrow band manganites, the observed CO modulation is 2-dimensional (2D). The absence of any extra HOLZ-rings in the CBED pattern taken along [100] zone of the CO phase at 100K further confirms the 2D nature of the observed CO modulation. The above described TEM observations thus clearly reveal that unlike stoichiometric $La_{0.2}Sr_{0.8}MnO_{3-\delta}$ ($\delta=0.01$) the off-stoichiometric $La_{0.2}Sr_{0.8}MnO_{3-\delta}$ ($\delta=0.12$) does not under go cubic to tetragonal phase-transition; rather transforms to a novel 2D CO phase.

The occurrence of cubic-tetragonal phase-transformation corresponding to the nearly stoichiometric $La_{0.2}Sr_{0.8}MnO_{3-\delta}$ can be clearly seen also in the XRD data shown in Fig.4 (a), (c). It shows splitting of (002) cubic peak into (004) and (220) peaks of I4/mcm tetragonal phase below $T_s=260K$ during cooling. But from Fig.4(d) it is clear that the off-stoichiometric sample does not show any such splitting. Fig.4(a) shows the temperature dependence of the lattice parameters of the cubic and tetragonal phases. The lattice parameters were obtained through Rietveld refinement [43] of the XRD data taken at various temperatures. Region between dashed lines represents phase-coexistence of Pm-3m and I4/mcm phases of the nearly stoichiometric $La_{0.2}Sr_{0.8}MnO_{3-\delta}$ ($\delta=0.01$). The tetragonality (c/a) of the basic perovskite distortion 80K is found to be 1.021. The elongation of c-axis indicates occupancy of $3d_{3z^2-r^2}$ orbitals all ferro-aligned along the c-axis resulting in cubic-tetragonal phase transformation. Relatively larger broadening of the (004) peak of I4/mcm phase as seen in Fig.4(c) is a direct evidence of the bulk occurrence of nano-twins, which are seen in the TEM micrograph in Fig.2(b) and (c). Fig.4 (a) shows monotonous decrease of the lattice parameter of the off-stoichiometric sample due to thermal contraction. Fig.4(b) shows variation of the specific-heat ($C_p$) with temperature during heating. A sharp transition is observed corresponding to the cubic-tetragonal phase-transition in the nearly stoichiometric sample. But this transition is completely absent for the off-stoichiometric sample, which finally confirms the absence of JT distortion induced cubic to tetragonal first order phase-transition in the off-stoichiometric sample. The occurrence of sharp transition in $C_p$-T measurement also indicates the high quality of the nearly stoichiometric $La_{0.2}Sr_{0.8}MnO_{3-\delta}$ ($\delta=0.01$) sample used in the present investigation.

$\rho$-T data of the nearly stoichiometric sample as well as off-stoichiometric sample are shown in Fig.5 (a). A step like feature in $\rho$-T at $T_s$= 260K followed by a monotonous increase can be seen for the nearly stoichiometric samples. The step feature shows a narrow but clear hysterisis indicating that it corresponds to the first order cubic to tetragonal phase-transition as seen in TEM and XRD results. The $\rho$-T of the off-stoichiometric sample does not show any such step-like feature but a monotonous increase of resistivity with decreasing temperature. It can be noticed that the room temperature resistivity values of the off-stoichiometric $La_{0.2}Sr_{0.8}MnO_{3-\delta}$ is ~4 times to that of the nearly stoichiometric one. To explore the possible transport mechanism operative in the case of off-stoichiometric sample nearest-neighbor hoping (log($\rho$)-vs-1/T), variable range hoping (VRH) (log($\rho$)-vs-$T^{-1/4}$) and log($\rho$)-vs-log(T) plots were made. These are shown in Figs.2(b) and (c). The nearest-neighbor hoping mechanism does not appear to be operative. But at first glance VRH and power-law both appear to fit to the $\rho$-T corresponding to the off-stoichiometric $La_{0.2}Sr_{0.8}MnO_{3-\delta}$ sample. After critical evaluation of the fits (solid-lines) it was found that VRH best fits down to ~120K and below 120K the power-law i.e. $\rho=C.T^{-\alpha}$ ($\alpha=9.5$) appears to best fit. There is a smooth cross-over from VRH to power-law transport. Power-law transport in manganites has been reported [44] in Cr-doped layered manganite at 250K. As will be discussed below both of the mechanisms, VRH and power-law, are expected in the case of disordered insulators.

Fig.6 shows 5T M-T data of the two $La_{0.2}Sr_{0.8}MnO_{3-\delta}$ samples. The sharp drop of magnetization at ~260K for the nearly stoichiometric $La_{0.2}Sr_{0.8}MnO_{3-\delta}$ corresponds to the C-type AFM phase-transition coupled with the cubic-tetragonal phase-transition as seen above through TEM and XRD investigations. The M-T data for nearly stoichiometric $La_{0.2}Sr_{0.8}MnO_{3-\delta}$ is identical to that of in Ref.27&33 and therefore approves the high quality of our nearly stoichiometric $La_{0.2}Sr_{0.8}MnO_{3-\delta}$ sample. It therefore also approves that the presently observed paramagnetic cubic to C-type AFM tetragonal phase-transition is the true behaviour of a stoichiometric ($\delta=0.00$) and nearly stoichiometric ($\delta=0.01$) $La_{0.2}Sr_{0.8}MnO_{3-\delta}$. On the other hand M-T data of the off-stoichiometric $La_{0.2}Sr_{0.8}MnO_{3-\delta}$ ($\delta=0.12$) sample does not show any sharp drop of magnetization. This indicates that C-type AFM transition is totally absent for the off-stoichiometric sample. M-T data of the off-stoichiometric $La_{0.2}Sr_{0.8}MnO_{3-\delta}$ does show a brad maxima ~ 215K, which may be due to increasing AFM interaction due to CO phase. The other note worthy feature of M-T is the higher value of magnetization for the off-stoichiometric sample down to 320K. These observations directly imply that oxygen off-stoichiometry plays an important role in stabilizing the type of magnetic interactions and related CO phase in these samples.

The above discussed TEM, XRD, $C_p$, $\rho$-T and M-T data clearly and independently approve that the nearly stoichiometric and off-stoichiometric $La_{0.2}Sr_{0.8}MnO_{3-\delta}$ samples bear characteristic differences between their structural and physical properties. The nearly stoichiometric $La_{0.2}Sr_{0.8}MnO_{3-\delta}$ under goes a first-order cubic to tetragonal phase-transition, with c/a =1.021. The tetragonality of c/a>1 is due to JT distortion resulting from ferro-ordering of $3d_{3z^2-r^2}$ orbitals [13]. This also gives rise to C-type antiferromagnetic ordering [9]. A comparison of the data obtained from nearly stoichiometric and off-stoichiometric $La_{0.2}Sr_{0.8}MnO_{3-\delta}$ samples clearly reveals that the JT distortion induced first order phase transition is totally absent for the off-stoichiometric sample. The present observation of a CO phase for the off-stoichiometric $La_{0.2}Sr_{0.8}MnO_{3-\delta}$ directly approves that a CO may stabilize even in the case of suppressed CJTD.

## DISCUSSIONS

Manganites are charge-transfer type insulators with strong coulomb repulsion (U). When coulomb repulsion is much stronger than the kinetic energy i.e. when U>>W, the $e_g$-electrons of $Mn^{+3}$ ions get localized in an ordered manner to minimize its total energy. Most of the time the ordered localization of $e_g$-electron is accompanied by CJTD and antiferromagnetic spin ordering which further minimize the total energy of the system ($E_{Tot}=U-E_{JT}-J_H$). In mixed-valent manganites the ordered localization (or disproportionation) of $e_g$-charge and the minimization of lattice-strain through CJTD [45,46] leads to Herring-Bone type arrangement of $3d_{3z^2-r^2}$ orbitals resulting in commensurate as well as incommensurate charge density modulation [4-6] with CE-type spin-ordering. The effective and important role of long-range coulomb repulsion in stabilizing charge-ordering has been experimentally realized by Radaelli et al [47] but yet there is no report of the occurrence of a CO phase in manganites which does not show CJTD and where coulomb-repulsion and exchange interaction can be attributed as the only factors to stabilize it. The currently observed LT-TEM and LT-XRD results for the off-stoichiometric $La_{0.2}Sr_{0.8}MnO_{3-\delta}$ clearly show the occurrence of CO superlattice spots without any systematic distortion of the $MnO_6$ octahedra [48] i.e. without CJTD or any other type of coherent JT distortion. In such a situation the occurrence of a CO phase directly implies that CO is being stabilized solely due to electron-correlation and Hunds coupling effect. In the present case of suppressed CJTD the CO will have dominantly exchange driven interactions [18,46,49].

In CMR manganites the absence of JT driven lattice-distortion has been attributed either to orbital-liquid [50] i.e. randomly positioned $(x^2-y^2)$, $(y^2-z^2)$ and $(z^2-x^2)$ orbitals or to the ordering of orbitals with complex coefficients, $|\pm\rangle = 1/\sqrt{2}[|z^2\rangle + i |x^2-y^2\rangle]$ i.e. a complex orbital ordering [20-22]. Since in the case of orbital-liquid model there is no translational order therefore a CO super-lattice will never occur according to this. Theoretically a complex orbital ordering demands formation of magnetic octapoles, which are supposed to bear very small magnetic moments [51]. Therefore the occurrence of charge-ordered superlattice spots and reasonable high magnetization values found in the present case discard the possibility of orbital-liquid and complex orbital ordered phases respectively. The suppression of JT distortion and differences observed in other physical properties of the off-stoichiometric sample can be understood as in the following.

Occurrence of oxygen off-stoichiometry will change the hole/electron or $Mn^{4+}/Mn^{3+}$ ratio of the sample with respect to the stoichiometric one. For the truly stoichiometric $La_{0.2}Sr_{0.8}MnO_{3-\delta}$ sample (i.e. $\delta=0.00$) the $Mn^{4+}/Mn^{3+}$ ratio will be 0.80/0.20 i.e. 4:1. With the decrease of each $O^{2-}$ ion two $Mn^{4+}$ ions will get converted into two $Mn^{3+}$ ions. Thus for the currently prepared nearly stoichiometric ($\delta=0.01$) sample $Mn^{4+}/Mn^{3+}$ will be 0.78/0.22 and for the off-stoichiometric sample with $\delta=0.12$ the $Mn^{4+}/Mn^{3+}$ ratio will be 0.56/0.44, which is rather close to half doped type i.e. x = 0.5. Therefore, based on the common knowledge about the La/Sr based manganites, the off-stoichiometric sample should mostly have been dominated by double-exchange driven ferromagnetic interaction and should have shown much lower resistivity as compare to that of the nearly stoichiometric one. On the contrary the studied off-stoichiometric sample shows ~ 4 times larger resistivity at room temperature, which further increases at lower temperatures with rate faster than that of the nearly stoichiometric one. This also transforms to a CO phase, that too with a modulation period of ~9 and not ~4 along [110]. Like half doped manganites [4,6] the estimated $Mn^{4+}/Mn^{3+}$ ratio of 0.56/0.44 for the studied off-stoichiometric sample a CO modulation period of only ~4 is expected. The observed CO modulation period of ~9 for the off-stoichiometric sample, is very close to the expected period of 10 for the truly stoichiometric $La_{0.2}Sr_{0.8}MnO_{3-\delta}$ ($\delta=0.00$).

The current observations thus clearly indicate that from CO point of view the off-stoichiometric sample is behaving like a stoichiometric sample. This suggests that the extra $Mn^{3+}$ ions, which appear after conversion from $Mn^{4+}$ due to off-stoichiometry in the sample, are not at all effective. As if the very source of its generation also acts as a sink too. This has been elaborated as in the following.

As described in the above the nearly stoichiometric $La_{0.2}Sr_{0.8}MnO_{3-\delta}$ ($\delta=0.01$) transforms to C type AFM tetragonal phase, in which c-axis expands and a,b axes shrink. C type AFM spin arrangement comprises c-axis oriented ferromagnetic spin-chains interacting anti-ferromagnetically in the a-b plane. Due to ferromagnetic interaction along the c-axis $e_g$-charge associated with $Mn^{3+}$ ions will maximizes its hopping along it by polarizing the charge density lobs of the $3d_{z^2-r^2}$ type orbitals along c-axis. This will result in ferro orbital ordering of $3d_{z^2-r^2}$ orbitals along c-axis. Due to this orbital ordering a coherent JT-distortion takes place causing expansion of c-axis and contraction of **a** and **b** axes resulting in cubic-tetragonal phase transformation. The off-stoichiometry will create oxygen-vacancy (OV). Recently Lin et al [52] through density function calculation on $SrTiO_3$ have shown that lowering of the local symmetry from cubic ($O_h$) to tetragonal ($C_{4v}$) around a OV in transition metal (TM) perovskite oxides induces direct on-site coupling between the $3d_{3z^2-r^2}$ and 4s,4p orbitals of the TM atom adjacent to the OV. Due to OV induced imbalance of direct coulomb interaction between $e_g$-charge and $O^{2-}$ ions the charge density lobs of $3d_{3z^2-r^2}$ orbitals of the TM atoms adjacent to OV, now get localized about the OV along the $O^{2-}$-TM-OV-TM-$O^{2-}$. Such a situation at one hand introduces a direct hopping between the two TM atoms lying just across the OV but on the other hand will suppress local JT-distortion around OV-site and also reduces indirect hopping between the next nearest TM atoms. For uniform oxygen deficiency of $\delta=0.12$ the estimated OV-OV distance will be ~8Å. Presence of large concentration of OV in the case of off-stoichiometric sample thus decreases the effectiveness of $Mn^{3+}$ ions towards its capability of orbital-ordering along c-axis and act as pinning centers against coherent JT distortion. Here it appears imperative to point out that unlike individual molecules the JT-effect in solids is realized as coherent distortion of local units known as "vibrons" [53], which in turn causes ferro distortive phase-transition. In $ABO_3$ perovskite oxide, containing B as JT active ion, the $BO_6$ octahedra represent the "vibron" units. It has been shown [54, 55] that for the coherent JT distortion a threshold concentration of JT-ions is necessary. Below threshold the activity of isolated JT-ions becomes ineffective. The present case of OV effectively leads to a similar situation. Consequently there will be no cubic-tetragonal phase-transformation in off-stoichiometric sample. Since the sites for the creation of OV will be randomly selected the orientation of $Mn^{3+}$-OV-$Mn^{3+}$ units will be equally probable along the three principle 4-fold axes X,Y,Z of the octahedral perovskite lattice and therefore the over all symmetry of the defect averaged long range phase will still remain cubic. Now since the dominating mechanism of tetragonal distortion is no more present the coulomb interaction (**U**) and Hunds coupling (**$J_H$**) may lead to a CO phase. Due to the presence of $Mn^{3+}$-OV-$Mn^{3+}$ like pinning sites the CJTD will also not take place. Due to hindrance from the large number of defect sites, the resulting CO phase will not be spatially very coherent. That is why the CO superlattice spots observed in the electron diffraction patterns in Fig.3 are rather diffuse.

Since the extra $e_g$-charge ($Mn^{3+}$) created due to off-stoichiometry will be localized around the OV, which also causes added scattering, the electrical resistivity of the off-stoichiometric sample increases in place of decreasing. Occurrence of large number of scattering centers in an insulator is known to cause Mott's VRH like transport. As shown in Fig.5(b) VRH is operative down to ~120K. Below ~120K the resistivity remains below the line dictated by VRH and it fits better with a power-law ($\rho=C.T^{-\alpha}$) behaviour. Occurrence of power-law transport in manganite has been reported earlier also. In Cr doped layered $La_{1.2}Sr_{1.8}Mn_{2-y}Cr_yO_7$ manganite a power-law transport with α=4/3 (1.33) [44] has been reported. It has been explained based on Glazman and Matveev (GM) model [56] and has been attributed to inelastic tunneling through channels with two impurities. In VRH process the hopping range of charge carriers increases with decreasing temperature. Therefore for sufficiently large sample size VRH will be applicable down to very low temperatures. The only limiting process to VRH is the tunneling of charge carries [56] through the localized states. Below certain temperature the hopping range will include more such states and the effective conduction will be dominated by the tunneling process. Below that temperature the resistivity of the sample will be lower than the VRH value. That is what has been observed in the present case. In tunneling conduction the temperature dependence of resistivity is dominated by inelastic, multistep tunneling via **n ≥ 2** impurities resulting in power-law temperature dependence with index **α = n-[2/(n+1)]** [44,56]. The channel with the least resistance decides the effective power-law. In off stoichiometric sample OV causes lowering of the local symmetry from $O_h$ to non-centrosymmetric $C_{4v}$ at the Mn-sites. This non-centrosymmetry will cause local spin canting between the Mn moments adjacent to the OV. The canting will give rise to finite ferromagnetic interaction between Mn moments and hence tunneling probability through $Mn^{3+}$-OV- $Mn^{3+}$ defect units will be more than rest of the defect-free nearby sites in the lattice. It appears that below 120K the hoping range is sufficiently large to include many such tunneling states and thereby resulting into a power-law transport. The observed value of α=9.5 indicates an effective inelastic tunneling through channels involving ~9-10 impurities in the channels.

As described in the above the local canted spins will bear extra moments above the paramagnetic background corresponding to nearly stoichiometric sample. That is why the off-stoichiometric sample bears higher magnetic moment than the nearly stoichiometric one at ~350K. With decreasing temperature the AFM interaction in the off-stoichiometric sample increases due to premonition of the CO phase and therefore its magnetization finally becomes lower below ~320K than that of the paramagnetic nearly stoichiometric sample. The nearly stoichiometric sample under goes strong C-type AFM phase transition with sharp drop in magnetization, where as the magnetization of the off-stoichiometric sample remains nearly flat showing weak ferromagnetic behaviour. This is due to saturation of the effective ferromagnetic interaction of the local canted moments at 5T field. The weak and broad maximum at ~215K is due to broad AFM transition of the CO phase.

## CONCLUSIONS

Based on above results and discussion it is concluded that in wide band manganite $La_{0.2}Sr_{0.8}MnO_{3-\delta}$ the charge ordered phase appears due to oxygen off-stoichiometry. The stoichiometric or nearly stoichiometric $La_{0.2}Sr_{0.8}MnO_{3-\delta}$ ($\delta=0.01$) under goes ferro ordering of $3d_{3z^2-r^2}$ orbitals causing coherent JT-distortion, which induces cubic to C-type AFM tetragonal phase transition. But in off-stoichiometric $La_{0.2}Sr_{0.8}MnO_{3-\delta}$ ($\delta=0.12$) the coherent JT-distortion gets suppressed due to localization of $e_g$ charge around the oxygen vacancy sites and therefore cubic to tetragonal phase transition does not take place. The important conclusion of the current observation is that a charge ordered phase can stabilizes even without CJTD. It appears that in the absence of coherent JT-distortion the Hunds coupling and coulomb repulsion now stabilize the charge ordered phase. The off-stoichiometric $La_{0.2}Sr_{0.8}MnO_{3-\delta}$ shows a power-law transport due to inelastic multistep tunneling through oxygen vacancy induced electronic states. Lowering of the local symmetry from $O_h$ to non-centrosymmetric $C_{4v}$ at the Mn-sites adjacent to oxygen vacancies causes formation of canted Mn moments, which results in weak ferromagnetic behaviour of the off-stoichiometric $La_{0.2}Sr_{0.8}MnO_{3-\delta}$.

**Acknowledgment**

Authors gratefully acknowledge Dr.P.Chaddah, Prof.A.Gupta and Dr.V.Ganesan for their encouragement and support. Aga Shahee would like to acknowledge CSIR-India for financial support as SRF.

**References:**

1. M. Imada, Atsushi Fujimori, and Yoshinori Tokura, Rev. Mod. Phys. **70**, 1039 (1998)
2. Y. Tokura, Rep. Prog. Phys. 69, 797 (2006)
3. D. I. Khomskii, Basic Aspects of the Quantum Theory, Cambridge University Press (2010)
4. C. H. Chen and S-W. Cheong Phys. Rev. Lett **76** 4042 (1996)
5. 4 S. Mori, C. H. Chen, and S-W. Cheong, Nature (London) **392**, 473 (1998)
6. J. C. Loudon,S. Cox,A. J. Williams,J. P. Attfield,P. B. Littlewood,P. A. Midgley,and N. D. Mathur ,Phys. Rev. Lett **94** 097202(2005)
7. G.C.Milward, M.J.Calderón and P.B. Littlewood Nature **433**, 607 (2005); S. Cox, J. Singleton, R. D. Mcdonal , A. Migliori and P. B. Littlewood Nature materials 7, 25 (2008)
8. A. Daoud-Aladine, J. Rodríguez-Carvajal, L. Pinsard-Gaudart, M. T. Fernández-Díaz, and A. Revcolevschi, Phys. Rev. Lett. **89**, 097205 (2002)
9. L. Wu, R. F. Klie, Y. Zhu, and Ch. Jooss, Phys. Rev. B **76**, 174210 (2007)
10. C. Zener, Phys. Rev. **82**, 403 (1951)
11. G.Subıas, J.Garcıa, J.Blasco, M.G. Proietti, H.Renevier, and M.C.Sanchez1, Phys.Rev.Lett. **93,** 156408 (2004)
12. J van den Brink, G. Khaliullin, and D. Khomskii, Phys.Rev.Lett. **83**, 5118 (1999)
13. R. Maezono, S. Ishihara, and N. Nagaosa, Phys. Rev. B **57**, R13993 (1998)
14. D. I. Khomskii, and K. I. Kugel, Phys. Rev. B **67**, 134401 (2003)
15. B. B. Van Aken, O. D. Jurchescu,A. Meetsma,Y. Tomioka,Y. Tokura, and Thomas T. M. Palstra Phys. Rev. Lett **90** 066403 (2003)
16. T. Hotta, Y. Takada, H. Koizumi and E. Dagotto, Phys.Rev.Lett. **84,** 2477 (2000)
17. H. Aliaga, D. Magnoux, A. Moreo, D. Poilblanc,2S. Yunoki,and E. Dagotto, Phys. Rev. B **68**, 104405 (2003)

18. J. B. Goodenough, Phys. Rev. **100**, 564 (1955)
19. E. Dagotto, New Journal of Physics **7** 67(2005)
20. D. Khomskii, arXiv: cond-mat/0004034v2 (2000)
21. R. Maezono, and N. Nagaosa, Phys. Rev. B **62**, 11576 (2000)
22. J van den Brink, and D. Khomskii, Phys. Rev. B **63**, 140416 (R) (2001)
23. H. Kawano, R. Kajimoto, M. Kubota, and H. Yoshizawa Phys. Rev. B **53** R14709 (1996)
24. J.S. Zhou and J. B. Goodenough, Phys. Rev. B 62, 3834 (2000)
25. J.S. Zhou and J. B. Goodenough, Phys. Rev. B **64**, 024421 (2001)
26. G.L. Liu, J.S. Zhou, and J. B. Goodenough, Phys. Rev. B **64**, 144414 (2001)
27. O. Chmaissem, B. Dabrowski, S. Kolesnik, J. Mais, J. D. Jorgensen, and S. Short, Phys. Rev. B **67**, 094431 (2003)
28. R. Maezono, S. Ishihara, and N. Nagaosa, Phys. Rev. B 58, 11583 (1998)
29. M. Izumi, Y. Konishi, T. Nishihara, S. Hayashi, M. Shinohara, M. Kawasaki, and Y. Tokura, Appl. Phys. Lett. **73**, 2497 (1998)
30. M. Izumi, T. Manako, Y. Konishi, M. Kawasaki, and Y. Tokura, Phys. Rev. B **61**, 12187 (2000)
31. K. Horiba, M. Taguchi, A. Chainani, Y. Takata, E. Ikenaga, D. Miwa, Y. Nishino, K. Kobayashi, T. Ishikawa, and S. Shin, Phys. Rev. Lett **93,** 236401 (2004)
32. S. I. Patil et al. Phys. Rev. B **62**, 9548 (2000)
33. J. Hemberger, A. Krimmel, T. Kurz, H. A. Krug von Nidda, V. Yu. Ivanov, A. A. Mukhin, A. M. Balbashov, and A. Loidl, Phys. Rev. B **66**, 094410 (200
34. H. Fujishiro, T. Fukase, and M. Ikebe, J. Phys. Soc. Jpn. **67**, 2582 (1998)
35. B. J. Sternlieb, J. P. Hill, U. C. Wildgruber, G. M. Luke, B. Nachumi, Y. Moritomo, and Y. Tokura, Phys. Rev. Lett. 76, 2169 (1996).
36. Y. Murakami, H. Kawada, H. Kawata, M. Tanaka, T. Arima, Y. Moritomo, and Y. Tokura, Phys. Rev. Lett. 80, 1932 (1998).
37. J. Q. Li, Y. Matsui, T. Kimura, and Y. Tokura, Phys. Rev. B 57, R3205 (1998).
38. T. Chatterji, G. J. McIntyre, W. Caliebe, R. Suryanarayanan, G. Dhalenne, and A. Revcolevschi, Phys. Rev. B 61, 570 (2000).
39. M. Merz, G. Roth, P. Reutler, B. Büchner, D. Arena, J. Dvorak, Y. U. Idzerda, S. Tokumitsu and S. Schuppler, Phys. Rev. B **74**, 184414 (2006).

40. L. J. Zeng, C. Ma, H. X. Yang, R. J. Xiao, J. Q. Li, and J. Jansen, Phys. Rev. B **77**, 024107 (2008).
41. R Bindu, Ganesh Adhikary, Nishaina Sahadev, N. P. Lalla, and Kalobaran Maiti, Phys. Rev. B **84**, 052407 (2011)
42. P.R. Sagdeo, Shahid Anwar, and N. P. Lalla, Phys. Rev. B **74**, 214118 (2006).
43. J. Rodríguez-Carvajal, Physica B **192**, 55 (1993)
44. K. B. Chashka, B. Fisher, J. Genossar, A. Keren, L. Patlagan, and G. M. Reisner, Phys. Rev. B **65** 134441(2002)
45. D. I. Khomskii, and K. I. Kugel Phys. Rev. B **67,**134401 (2007)
46. K. I. Kugel, and D. I. Khomskil, Zh. Eksp. Teor. Fiz. **64**, 1429 (1973)
47. P. G. Radaelli, D. E. Cox, L. Capogna, S.-W. Cheong, and M. Marezio, Phys. Rev. B **59**, 14440 (1999)
48. T. Nagai, T. Kimura, A. Yamazaki, T. Asaka, K. Kimoto, Y. Tokura, and Y. Matsui1, Phys. Rev. B **65**, 060405(R) (2002).
49. D. I. Khomskii, Physica Scripta, **72**, CC8 (2005)
50. S. Ishihara, M. Yamanaka, and N. Nagaosa, Phys. Rev. B **56**, 686 (1997)
51. P. Santini and G. Amoretti Phys. Rev. Lett **85** 2188 (2000)
52. C. Lin, C. Mitra, and A. A. Demkov, Phys. Rev. B **86** (2012) 161102(R)
53. G. A**.** Gehring, and K. A**.** Gehring, Rep. Prog. Phys. **38,** 1 (1975)
54. Donald S. Mcclure, J.Phys. Chem. Solids **3**, 311(1957)
55. J. D. Dunitz, and L. E. Orgel, J. Phys. Chem. Solids **3**, 318 (1957)
56. L. I. Glazman and K. A. Matveev, Sov. Phys. JETP **67** 1276 (1988)

**Figure captions:**

**Fig.1.** Rietveld refined XRD patterns of nearly stoichiometric and off-stoichiometric $La_{0.2}Sr_{0.8}MnO_{3-\delta}$ samples. Absence of any unaccounted peak confirms that both the samples are pure cubic phase with space group Pm-3m.

**Fig.2.** Electron micrographs taken from the same area (a) at 300K and (b) at 100K. (c) shows an enlarged view of a nano-twined (10-20 nm) microstructure of the tetragonal phase appearing after transformation below 260K.(d)&(e) show SAD patterns from the same region in its cubic and tetragonal states respectively. The splitting of the spots in (e) indicates (011) mirror relation between twins.

**Fig.3.** (a) High-resolution micrograph taken from the charged-ordered (CO) phase observed for off-stoichiometric $La_{0.2}Sr_{0.8}MnO_{3-\delta}$ at 100K and (b) The corresponding SAD pattern of CO superlattice spots showing ~ 9x9 modulations along [011] and [0-11]. It can be noticed that the observed CO lattice fringes are not straight. It depicts that the **$e_g$**-charge is not strongly pinned with the lattice due to lack of cooperative JT-distortion. (c) The intensifying of CO superlattice spots as a function of cooling can be noticed

**Fig.4.** (a) lattice-parameter variation with temperature of the nearly stoichiometric and off-stoichiometric $La_{0.2}Sr_{0.8}MnO_{3-\delta}$ samples. (b) $C_p$-T data for both of the $La_{0.2}Sr_{0.8}MnO_{3-\delta}$ samples. It shows complete absence of the δ-transition for off-stoichiometric $La_{0.2}Sr_{0.8}MnO_{3-\delta}$ samples. (c)&(d) show temperature dependent XRD profiles of the (200) peaks of the two $La_{0.2}Sr_{0.8}MnO_{3-\delta}$ samples during cooling down to 80K. In (c) the splitting of (200) XRD peak of cubic phase into (004) and (220) of the tetragonal (c/a=1.021) transition of the nearly stoichiometric sample. But no such splitting is observed for the off-stoichiometric sample in (d).

**Fig.5.** (a) Semi-log plots of resistivity (ρ) - vs - temperature (T) data corresponding to nearly stoichiometric (δ=0.01) and off-stoichiometric (δ=0.12) $La_{0.2}Sr_{0.8}MnO_{3-\delta}$ samples. (b) Nearest-neighbor hopping {ρ=A.exp(B/T)} and variable-range hopping (VRH) {ρ =A.exp(B/$T^{1/4}$)} plots of ρ(T)-T data corresponding off-stoichiometric sample. Deviation from linearity, as compared to the solid-line, indicates that VRH is effective only down to 120K. (c) log-log plot of ρ(T)-T data of off-stoichiometric sample. Linearity of the log-log plot indicates that below 120K power-law (i.e. R=C.$T^{-\alpha}$) type mechanism is operative in the off-stoichiometric $La_{0.2}Sr_{0.8}MnO_{3-\delta}$ sample.

**Fig.6.** 5T field-cool M-T data for the nearly stoichiometric (δ=0.01) and off- stoichiometric (δ=0.12) $La_{0.2}Sr_{0.8}MnO_{3-\delta}$ samples. Complete absence of C-type AFM transition at ~260K can be seen for the off- stoichiometric sample.

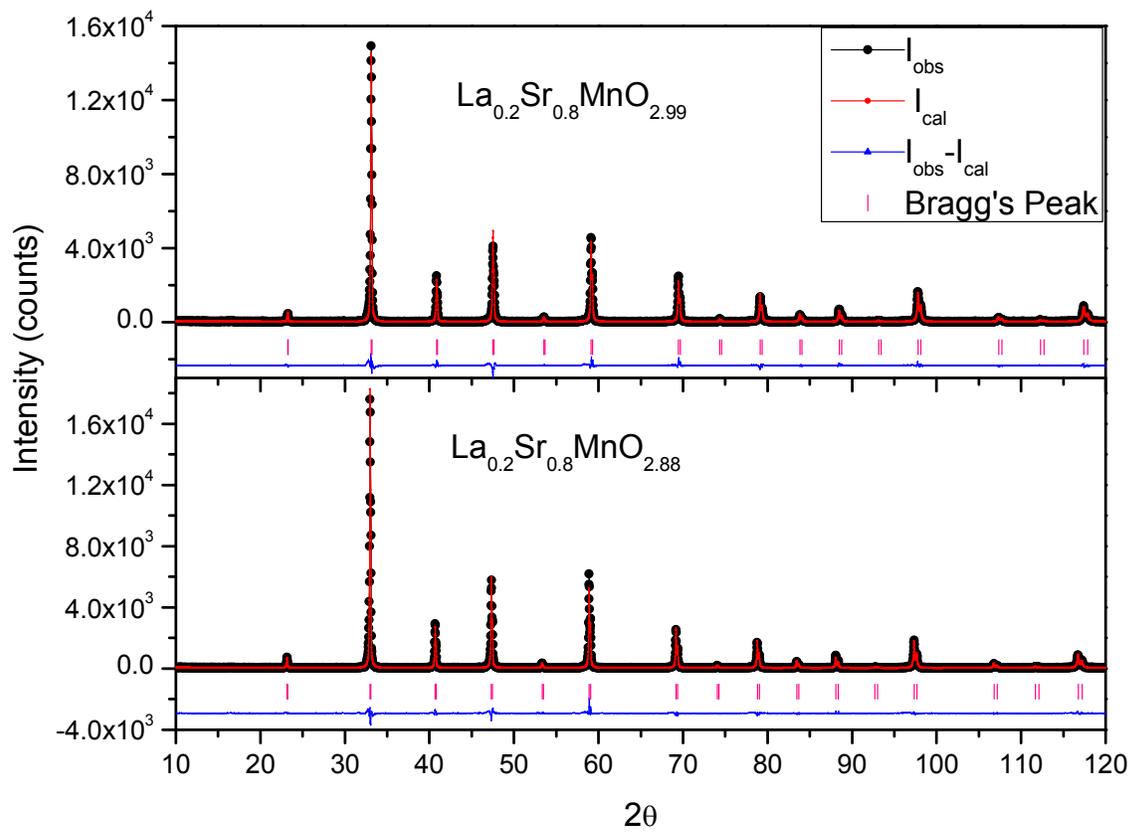

**Figure 1**

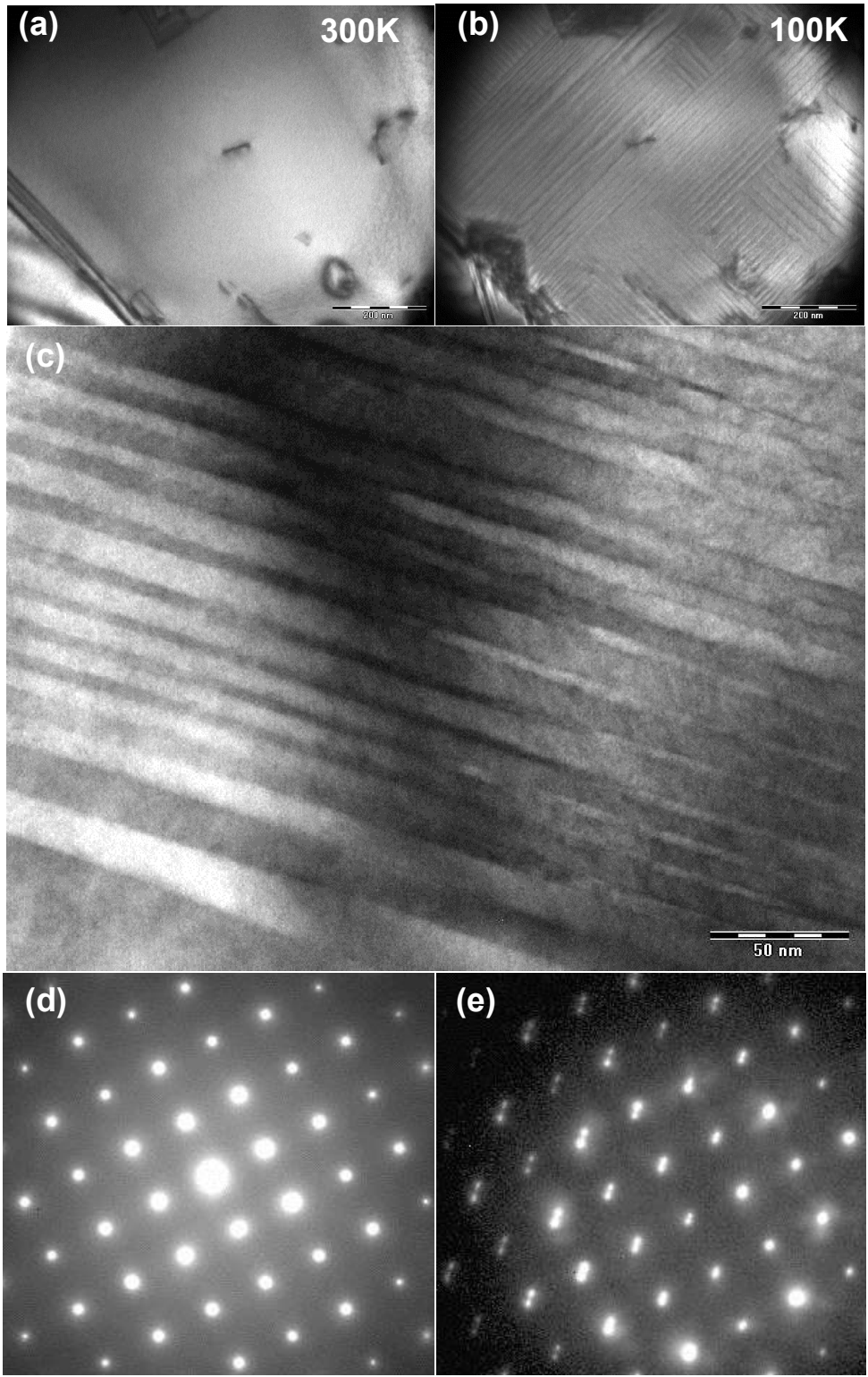

**Figure 2**

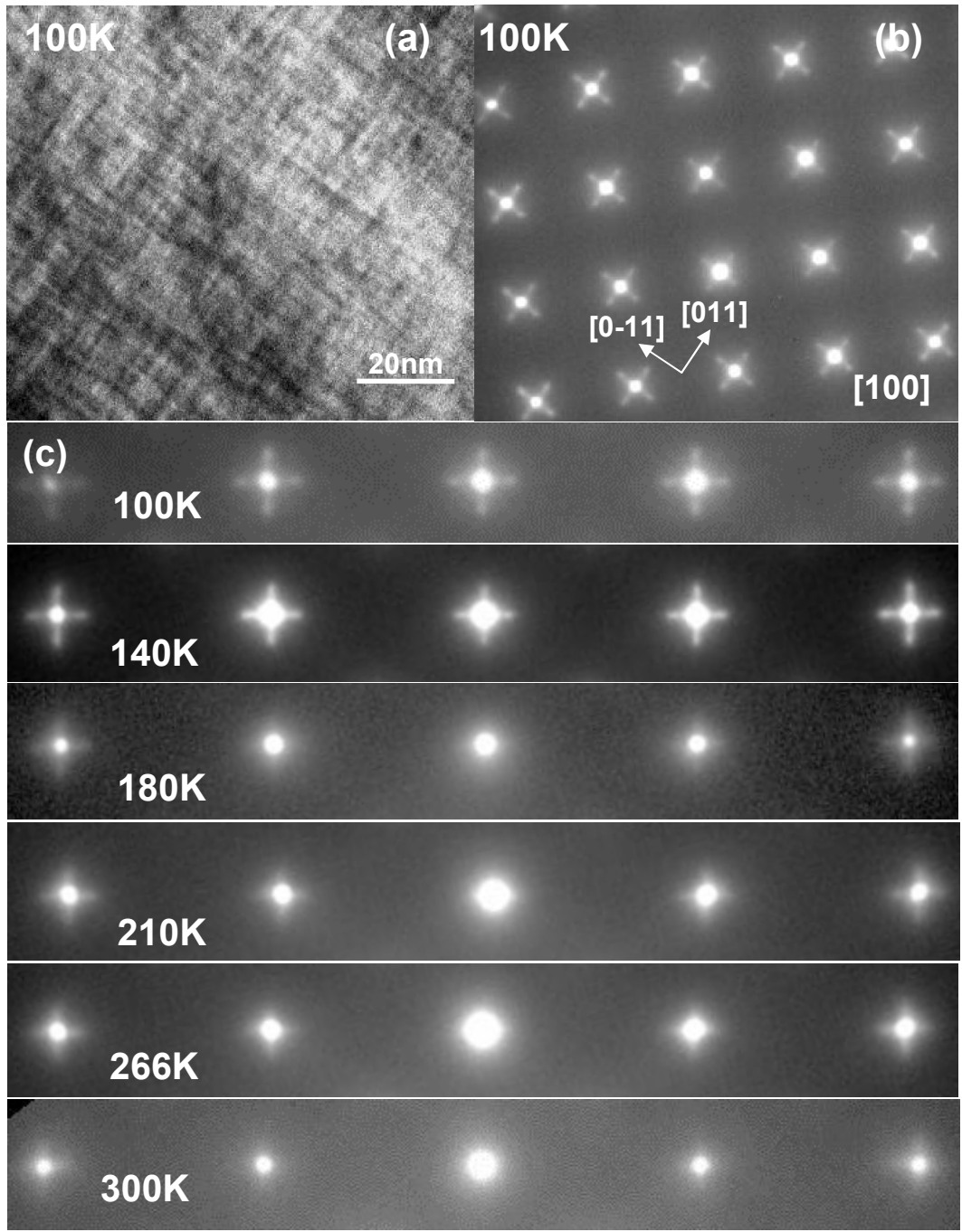

**Figure 3**

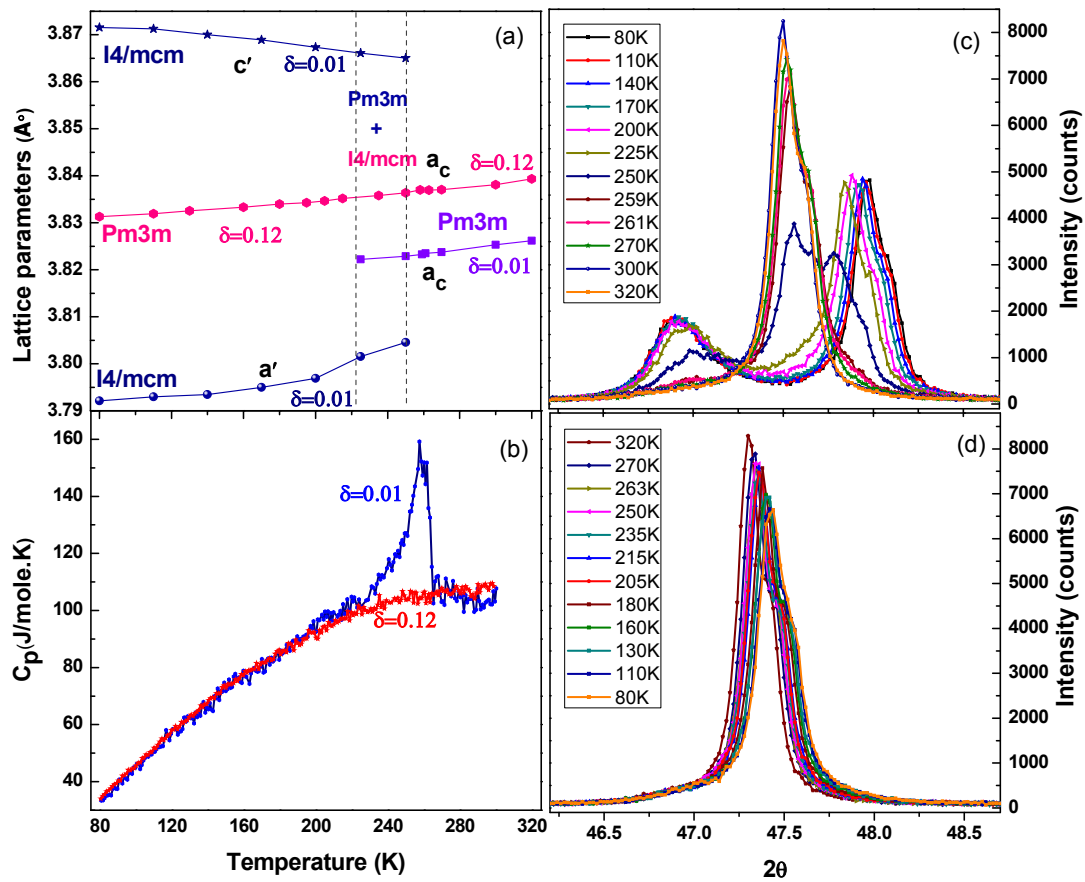

**Figure 4**

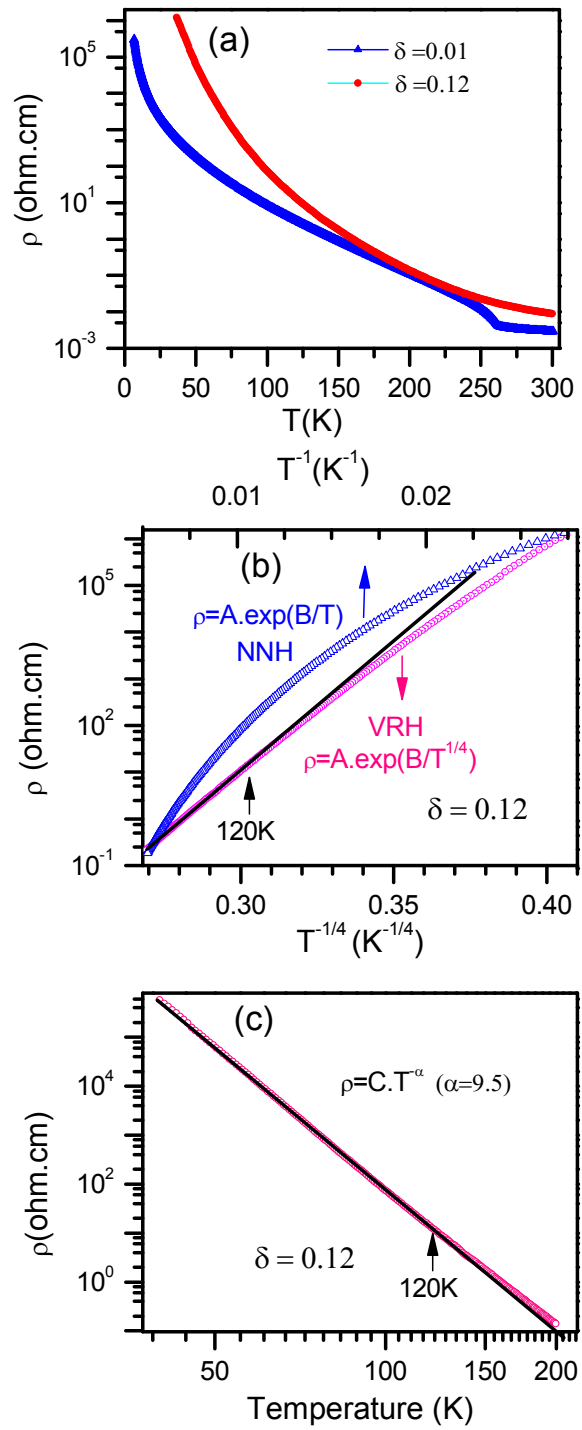

**Figure 5**

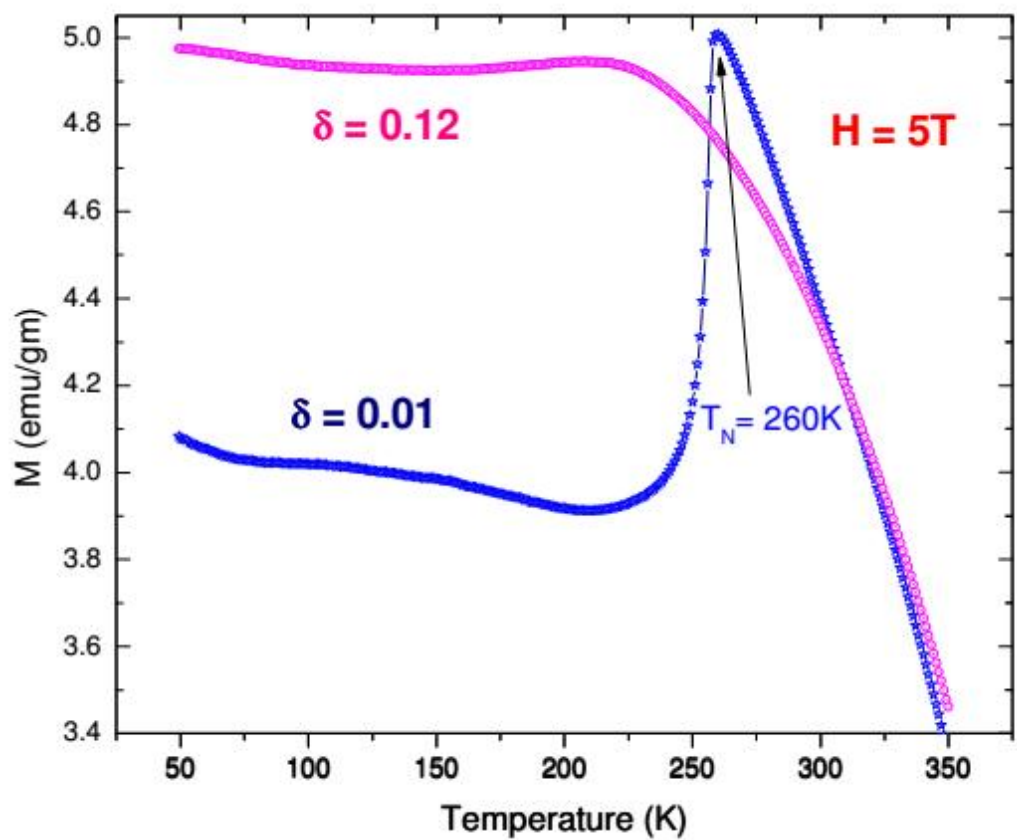

**Figure 6**